\documentclass[aps,showpacs,twocolumn,prl,preprintnumbers,nofootinbib]{revtex4}

\usepackage{bbm}
\usepackage{bm}

\newcommand{\nc}{\newcommand}
\nc{\beq}{\begin{equation}}  \nc{\eeq}{\end{equation}}
\nc{\bea}{\begin{eqnarray}}  \nc{\eea}{\end{eqnarray}}

\def\vev{vacuum expectation value}
\def\su#1{{SU(#1)}}
\def\ui{U(1)}

\def\lcal{{\cal L}}
\def\pcal{{\cal P}}
\def\qcal{{\cal Q}}
\def\mBB{{\mathbbm M}}
\def\rBB{{\mathbbm R}}
\def\vBB{{\mathbbm V}}
\def\mati{{\mathbbm1}}
\def\gev{\hbox{GeV}}
\def\tev{\hbox{TeV}}
\def\ssb{spontaneous symmetry breaking}
\def\alpbf{{\bm\alpha}}
\def\betbf{{\bm\beta}}
\def\gambf{{\bm\gamma}}
\def\delbf{{\bm\delta}}
\def\tr{ \hbox{tr}}
\def\HH{{\bf H}}
\def\xx{{\bf x}}
\def\yy{{\bf y}}
\def\imax{I_{\rm max}}
\def\bra#1{\left|#1\right\rangle}
\def\half{\frac12}
\def\ie{{\it i.e.}}

\begin{document}

\preprint{IFT-06-07\cr
UCRHEP-T411}

\title{5-Dimensional Difficulties of Gauge-Higgs Unifications}

\author{Bohdan GRZADKOWSKI}
\email{bohdan.grzadkowski@fuw.edu.pl}
\affiliation{Institute of Theoretical Physics,  Warsaw University,
Ho\.za 69, PL-00-681 Warsaw, Poland}

\author{Jos\'e WUDKA}
\email{jose.wudka@ucr.edu}
\affiliation{Department of Physics, University of California,
Riverside CA 92521-0413, USA}

\begin{abstract}
We consider 5 dimensional gauge theories where the 5th direction 
is compactified on the orbifold $S^1/Z_2$, and where
the 5th components of the gauge bosons play the role of 
the Standard Model Higgs boson (gauge-Higgs unification).
The gauge symmetry breaking is realized through the appropriate orbifold
boundary conditions and through the Hosotani mechanism. 
We show that for {\em any such theory} the assumption that the low-energy vector-boson 
spectrum consists of the $W^\pm$, $Z$ and $\gamma$ only, is inconsistent 
with the experimental requirements $\sin^2\theta_W\simeq 1/4$ and $\rho\equiv 
m_W^2/(m_Z^2\cos^2\theta_w)=1$. 
\end{abstract}

\pacs{11.10.Kk, 11.15.-q, 12.10.-g}
\keywords{gauge theories, Higgs boson, extra dimensions, gauge-Higgs unification}

\maketitle

{\it Introduction}
In the Standard Model (SM), the Higgs mechanism is responsible for generating
fermion and vector-boson masses. 
Although the model is renormalizable
and unitary, it has severe naturality problems
associated with the so-called ``hierarchy problem''.
At loop-level this problem reduces to the fact 
that the quadratic corrections tend to 
increase the Higgs boson mass up to the UV cutoff of the theory.
Extra dimensional extensions of the SM offer a novel
approach to gauge symmetry breaking in which the hierarchy problem could be
either solved or at least reformulated in terms of the geometry of the higher-dimensional space.
A particularly attractive scenario is offered by the Hosotani mechanism \cite{Hosotani:1988bm} 
where gauge symmetry breaking is generated by the \vev\ of
the extra component of the gauge field, $A_4$, whose Kaluza-Klein zero mode
plays a role of the 4-dimensional Higgs boson; a setup known as gauge-Higgs unification (GHU). 
Though 5-dimensional gauge symmetry and locality
prevent a tree-level potential for $A_4$, 
radiative effects generate a non-trivial effective potential,
leading to a prediction for the Higgs boson mass and the scale of gauge symmetry 
breaking. In such models the boundary conditions determine the gauge group
of the light sector (presumably $\su3 \times \su2 \times \ui$), and the  \vev\ $\langle A_4\rangle$
provides a second stage of breaking, presumably to $\ui_{EM}$.
The fact that the symmetry breaking pattern of  the low-energy theory is predicted
by the gauge and fermion~\footnote{Fermions enter the effective potential for the
zero mode of $A_4$ at a loop level.} structure of the fundamental theory 
is indeed very appealing. 
Other inherent problems of the SM could also be addressed in
extra-dimensional scenarios. For instance, within the SM
the amount of CP violation is not sufficient to explain the observed barion 
asymmetry~\cite{Barr:1979ye}, the GHU scenario offers a possible solution
since in such models the geometry can be a new source
of explicit and spontaneous CP violation~\cite{Grzadkowski:2004jv}. 

The most economic realization of the GHU paradigm uses $\su3_c\times\su3_w$
as the gauge group of the full theory~\cite{Antoniadis:1993jp}, 
however the model predicts the phenomenologically unacceptable value
of the weak-mixing angle~\footnote{We define
$\theta_W$ as the angle that diagonalizes the $Z$-$\gamma$ 
mass matrix.} $\theta_W=\pi/3$~\cite{Scrucca:2003ra}. 
Though there exist various remedies to this problem (localized gauge kinetic 
terms~\cite{Burdman:2002se}, \cite{Agashe:2004rs} or allowing a low-energy
gauge group with an extra -- anomalous -- $U(1)$~\cite{Antoniadis:2001cv},
\cite{Scrucca:2003ra}), in this paper we will restrict ourselves to the simplest (hence
more attractive) scenario and we will not pursue such options.
Other models also have serious problems, for example, when the gauge group is $SU(5)$
it is natural to expect spontaneous breaking of $SU(3)_c$~\cite{Haba:2002py}. 
The next minimal choice, $SU(6)$ again
suffers from the presence of an extra light $U(1)$
that must be broken by an extra elementary Higgs field~\cite{Haba:2004qf}.
So, in the simplest 5D examples of the GHU either $\sin^2\theta_W$ is not
phenomenologically acceptable, or the low energy gauge group is larger than 
$\su3 \times \su2 \times \ui$.

Because of these observations it is natural to ask whether there exist 5D 
GHU models with all fields propagating in the bulk without localized excitaitons and
where the light sector is an $\su3 \times \su2 \times \ui$ gauge theory
broken to $\ui$ by the Hosotani mechanism, and such that the predictions for
the weak-mixing angle and the oblique parameters 
are close to the experimental values. We will argue below that these constraints
cannot be satisfied; {\em no} such model is phenomenologically viable.

{\it The models}
The Lagrangian is assumed to have the form
$
\lcal = - \left( F_{MN}^a\right)^2/4\; + $ fermion, ghost and gauge-fixing terms,
where
$ F_{M N}^a = \partial_M A_N^a - \partial_N A_M^a  + g_5 f_{abc} A_M^b A_N^c $
(with $f_{abc}$ gauge-group structure constants and
$ g_5\sim({\rm mass})^{-1/2}$ the gauge coupling) and
$M,~N,\ldots=(0,1,2,3,4)$ the 5-dimensional space-time indices 
with the first four corresponding to Minkowski space
(labeled by Greek letters $\mu,~\nu, \ldots$). The
last index corresponds to the compact direction; we use $ x^4=y$.

We will consider a space of the form
$ \mBB \otimes (\rBB/\qcal)$ where $ \mBB$ denotes
the 4-dimensional Minkowski space-time and
$\qcal$ is a discrete group with two elements:
{\it(i)} Translation, $ y \to y+L $, where $L$ is
the size of the compact subspace; and {\it(ii)} reflection,
$ y \to -y $. 

We assume that under $\qcal$ the gauge fields 
transform according to~\cite{Grzadkowski:2005rz}
\beq
A_N^a(y+L)  = \vBB_{ab} A_N^b(y), \;
A_N^a(-y)  = (-1)^{\delta_{N,4}}\tilde\vBB_{ab} A_N^b(y),
\label{eq:vvt}
\eeq
where $ \vBB, ~\tilde\vBB$ are real and orthogonal matrices (in a basis where
the structure constants are real) representing involutions of the gauge algebra.
Note that the orbifolding (\ref{eq:vvt}) allows also for the generalized
twisting discussed in~\cite{Grzadkowski:2005rz}.
We will first assume that the gauge group is simple and then generalize.

For a simple group the
transformations (\ref{eq:vvt}) leave the Lagrangian invariant provided
\beq
\vBB_{d a} \vBB_{e b} \vBB_{f c} f_{ d e f} = f_{abc} ; \quad
\tilde\vBB_{d a} \tilde\vBB_{e b} \tilde\vBB_{f c} f_{ d e f} = f_{abc}.
\label{star}
\eeq

In addition (\ref{eq:vvt}) must provide a representation
of $ \qcal $. Using the fact
that $ -y = [-(y+L)]+L $ and that $ -(-y) = y$ we find 
\beq
\vBB\tilde\vBB\vBB = \tilde\vBB; 
\quad
\tilde\vBB^2 = \mati .
\label{eq:qcons}
\eeq

The models we consider are then defined by the
Lagrangian $ \lcal$, which specifies the dynamics, as well
as by the matrices $ \vBB, ~ \tilde\vBB $
that determine the behavior under $ \qcal $. 
Similar matrices are associated with the transformation
rules for the fermions~\cite{Grzadkowski:2005rz}, however those will not be relevant for the
arguments presented hereafter. 

{\it Light spectrum}
Higher-dimensional theories must satisfy the minimum
constraint of generating the experimentally observed light spectrum; because
of this it is of interest to derive the general properties of these
excitations. To this end it proves
convenient to expand the various fields in Fourier modes in the
compact coordinate $y$, the coefficients are then 4-dimensional  fields
for which the action of $ \partial_y$ generates a mass term; 
all $y$-dependent modes will then be heavy (mass $\sim1/L$) while
light excitations are associated with $y$-independent modes.

The light gauge bosons will be denoted by $A_\mu^{\hat a}$;
the light modes associated with $A_{N=4}$ behave
as 4-dimensional scalars and will be denoted by $ \phi_{\hat r} =
A^{\hat r}_{N=4} $. Using the $y-$independence of these modes and
the behavior of the field under $ \qcal $ we find~\cite{Grzadkowski:2005rz}
\bea
A_\mu^{\hat a} &=& \vBB_{\hat a \hat b} A_\mu^{\hat b} = 
\tilde \vBB_{\hat a \hat b} A_\mu^{\hat b},\cr
\phi^{\hat r} &=& \vBB_{\hat r \hat s} \phi^{\hat s}= 
-\tilde \vBB_{\hat r \hat s} \phi^{\hat s}.
\label{eq:light}
\eea
If we denote by
$P^+$ the subspace of generators characterized by $+1$
eigenvalues of $\tilde\vBB$ and $\vBB$ and
$N^+$ the subspace of generators characterized by $-1$
eigenvalues of $\tilde\vBB$, and $+1$ eigenvalues of $\vBB$,
then the light gauge bosons and scalars are associated with 
$P^+$ and  $N^+$, respectively. Denoting by $R$ the set of remaining 
generators we find that (\ref{star}) and  (\ref{eq:qcons}) imply that
\beq
\left[ N^+ , P^+\right] \subset N^+, \;
\left[ N^+ , N^+\right] \subset P^+, \;
\left[ N^+ , R \right] \subset R.
\label{nnp}
\eeq

Extracting from $ \lcal$ the terms that contain only light fields, we
find the usual gauge terms for the $ A^{\hat a}$ and the gauge-invariant
(under the subgroup associated with the $ A^{\hat a}$) kinetic terms for 
the $ \phi $. Note however that the form of $ \lcal$ 
disallows any tree-level potential for
$ \phi$; it follows that {\em at tree-level} all 4-dimensional 
bosons are either massless or have a mass $\sim1/L$.

If these models are to be phenomenologically
viable, they must be able to generate masses for 
the appropriate vector bosons at a characteristic scale $v \sim 100\gev$.
This symmetry breaking step can
result from  radiative corrections since these will
generate a non-vanishing (effective) potential $V_{\rm eff}$  for the $\phi$
at $ \ge1$ loops. This opens the possibility that these models will undergo
two stages of symmetry breaking: the first generated by the behavior 
under $ \qcal $ and the second, at a presumably lower scale, generated
radiatively by the scalars $\phi$.
Since the scale of $V_{\rm eff}$ is $1/L$ most
models predict both scalar and vector boson masses of $O(1/L)$,
in particular the $m_\phi$ is too light. This problem can find a natural solution by
choosing the gauge group, boundary conditions and fermion
content~\cite{Scrucca:2003ra},\cite{Haba:2004qf};
obtaining such a realistic symmetry breaking pattern is a
fundamental issue in the GHU scenario.

{\it Phenomenological constraints}
Here we consider those 5-dimensional models which contain only
gauge boson fields and whose light excitations are
described by an $\su3\times\su2\times\ui$ gauge group.
We assume that the $\phi$ effective potential will
lead to the expected pattern of \ssb; in addition
we require
\beq
\sin^2\theta_W \sim 0.25 ; \quad \rho \simeq 1
\label{phencon}
\eeq
at tree level.
Since we will exhibit a serious problem associated with 
the minimal requirements (\ref{phencon}) we will not investigate 
whether there exist models where the scalar effective
potential produces the correct pattern of \ssb. It is possible that no
such model exists, in which case our arguments can only be
strengthened. We then assume that the zero modes of the $A_4$ acquire \vev s
$ v \ll 1/L$ from an effective potential generated at one loop.

We denote by $E_\alpbf$ and $H_i$ the roots and Cartan generators
of Lie algebra of the full theory normalized such that
$ \tr H_i H_j = \delta_{ij}, ~\tr E_{-\betbf}E_\alpbf = \delta_{\alpbf,\betbf}$.
Then it  is straightforward to show that the generators of any $\su2$ subgroup
(a possible choice for the SM $SU(2)$) will be of the form~\cite{georgi}
\bea
J_0 = \frac{1}{| \alpbf|^2}  \alpbf \cdot \HH, \quad
J_+ = \frac{\sqrt{2} }{ | \alpbf| } E_{\alpbf}, \quad
J_-=(J_+)^\dagger.
\label{su2gen}
\eea
The SM hypercharge generator $Y$  generates a $\ui$ subgroup and
commutes with $J_{0,\pm}$, we then have
\beq
Y = \hat\yy\cdot\HH; \quad \hat\yy \cdot\alpbf=0.
\label{u1gen}
\eeq

The light scalars that can contribute to the
vector-boson mass matrix (in case they acquire a \vev)
can be arranged according to their $\su2 $ representations.
Assume first that one scalar containing light modes 
is associated with a linear combination of
Cartan generators $ \xx\cdot\HH$. Then, for
any root vector $\gambf$ such that $ \xx\cdot\gambf \not=0 $ we have
$ [\xx\cdot\HH , E_\gambf] = (\xx\cdot\gambf) E_\gambf$ 
which is consistent with (\ref{nnp})
only if $E_\gambf \in R$; in particular this implies that
$ \xx\cdot\HH$ commutes
with all generators associated with the SM $\su2\times U(1)$,
so the associated scalar
will be a singlet and cannot contribute to the mass structure of the light
vector bosons. 

Therefore, the light scalar state which is an eigenvector of $J_0$ with the eigenvalue $I$ that belongs
to a multiplet of isospin $\imax (\imax + 1)$ should be of the form~\footnote{In the
adjoint representation we will identify a state $\left| X_{\rm a} \right\rangle $
with a generator $X_{\rm a} $; the action of a generator
on such a state is given by 
$X_{\rm b} \left| X_{\rm a} \right\rangle = 
\left| \left[ X_{\rm b} , X_{\rm a} \right] \right\rangle $.}
\beq
\bra I = \sum_\betbf v^\betbf \bra{E_\betbf},
\label{state}
\eeq
then $ J_0 \bra I = I \bra I $ implies
\beq
\alpbf \cdot \betbf = | \alpbf|^2 I.
\label{abi}
\eeq
Note that $\betbf$ cannot be parallel to $\alpbf$.

Next we consider the repeated application of the lowering and rising
operators to $ \bra I $,
\beq
J_\pm^n \bra I = \sum_\betbf v^\betbf_{\pm n} 
\bra{E_{\betbf \pm n\alpbf } } .
\eeq
Note that not all such states will vanish (otherwise
$ \bra I$ would be an $\su2$ singlet), hence
$ E_{\betbf \pm n\alpbf }  \in N^+$ for some
integers $n$. Using (\ref{nnp}) we then find
\beq
\left[ E_{\betbf \pm n \alpbf } , 
E_{-\betbf \mp n \alpbf } \right]
= (\betbf \pm n \alpbf ) \cdot \HH \in P^+ , 
\label{ebcom}
\eeq
which implies that the set of generators
$ \{ (\betbf \pm n \alpbf ) \cdot \HH \}$, such that 
$ \betbf \pm n \alpbf $ is a root, are in $P^+$.

Suppose first that there are two root vectors
$\betbf,~\betbf'$ that contribute
to the sum (\ref{state}), then $P^+$ will contain generators
proportional to $(\betbf + n \alpbf ) \cdot \HH 
,~(\betbf' + n' \alpbf ) \cdot \HH $ 
(for some  integers $n,~n'$); in addition,
$P^+$ will also  contain $J_0$. But this is
impossible since the electroweak group has rank 2.
It then follows that a single root-vector $\betbf$
can contribute to the sum in (\ref{state}): the 
constraint on the rank allows a single light scalar
multiplet.

This also implies that the hypercharge generator
(\ref{u1gen}) must be of the form 
$ ( r \alpbf + s \betbf)\cdot\HH$ for some constants $r$ and $s$;
using (\ref{u1gen}) then implies
\beq
\hat\yy = \frac{ \betbf - ( \hat\alpbf \cdot \betbf) \hat\alpbf}
{ \left| \betbf - ( \hat\alpbf \cdot \betbf) \hat\alpbf \right|}.
\label{defofyhat}
\eeq

Then the (canonically normalized) electroweak bosons correspond to the zero modes
of the gauge fields associated with the generators
$ \hat\alpbf \cdot \HH, ~ E_{\pm\alpbf},~ \hat\yy \cdot \HH $;
we denote these zero modes by $W^0 , W^\pm$ and $B$
respectively.
Then we can write
\bea
A_\mu &=& W_\mu^+ E_\alpbf + W_\mu^- E_{-\alpbf} + W_\mu^0 \hat\alpbf\cdot\HH + B_\mu \hat\yy\cdot\HH
 + \cdots \cr
A_4 &=& \phi E_\betbf + \phi^\star E_{-\betbf} .
\eea

The terms in the Lagrangian responsible for the generation of vector bosons masses
are $ \propto \tr [ A_\mu ,A_4]^2 $. Using $ [E_\gambf, E_\delbf] = 
N_{\gambf , \delbf } E_{\gambf + \delbf} $
and the standard properties of the $ N_{\gambf , \delbf }  $~\cite{georgi}, we find
\bea
\tr[A_\mu,A_4]^2 &=&
\left(N_{\alpbf,\betbf}^2 + N_{-\alpbf,\betbf}^2\right) W^+\cdot W^- \cr &&
+\left[(\hat\alpbf\cdot\betbf) W^0 +  (\hat\yy\cdot\betbf)B\right]^2 + \cdots .
\eea

Now, $ N_{\gambf,\delbf}^2 = p (\gambf\cdot\delbf) + |\gambf|^2 p( p+1 )/2 $,
where $p$ is an integer such that $ p \gambf + \delbf$ is a root, but $ (p+1)\gambf + \delbf$
is not. For our case, using (\ref{abi}), we have $ p = \imax \mp I $ for
$ N_{\pm \alpbf,\betbf} $ so 
\beq
N_{\alpbf,\betbf}^2+
N_{-\alpbf,\betbf}^2 = |\alpbf|^2 \left[\imax(\imax+1) -I^2\right].
\eeq

Assuming that $\bra I $ is a member of a multiplet with
maximum isospin $ \imax $ and it is the component that gets a \vev\
$v/\sqrt{2}$, it is straightforward to show that the mass-terms
in $\lcal$ take the form 
\bea
\lcal_{\rm mass} &=& \frac{v^2}{2} \left\{
|\alpbf|^2 [ \imax (\imax+1) - I^2] W^+ \cdot W^- \right. \cr && \qquad \left.
+( \hat\alpbf \cdot \betbf W^0+ \hat\yy\cdot\betbf B)^2 \right\},
\label{lmass}
\eea 
so the electroweak mixing angle and $\rho$ parameter are given by
\beq
\sin^2\theta_W = 1 - ( \hat\alpbf \cdot \hat \betbf )^2, \quad
\rho = \frac{ \imax(\imax+1)}{2 I^2} - \half .
\eeq

Since either $\betbf + \alpbf$ or $\betbf - \alpbf$ is a root
(otherwise $I=\imax=0 $), then the commutator
$ [ E_\betbf , E_{\betbf \pm\alpbf } ] $ either vanishes
or it is proportional to $E_{2\betbf \pm\alpbf} $.
But since $E_\betbf,~ E_{\betbf \pm\alpbf } $ are roots, then,
using (\ref{nnp}) shows that a non-zero commutator
implies that $E_{2\betbf \pm\alpbf} = E_\alpbf $ or
$E_{2\betbf \pm\alpbf }= E_{-\alpbf} $ both of which are
impossible. Hence
\beq
 [ E_\betbf , E_{\betbf \pm\alpbf } ] =0 
\label{vancom}
\eeq

There are then two possibilities:

{\it(i)} $-\betbf \pm \alpbf$ is not a root. Then
$[ E_{-\betbf} , E_{\pm\alpbf } ] =0 $ which, together
with (\ref{vancom}) and
$[ E_{-\betbf} , E_{\betbf\pm\alpbf}] \propto E_{\pm\alpbf}$
imply
$ \betbf \cdot( \betbf \pm\alpbf)  = |\betbf|^2/2$.
Combining this with (\ref{abi}) we find
\beq 
|I|/2 = (\hat\alpbf\cdot\hat\betbf)^2 = m/4,
\eeq
where $m$ is an integer, $ 0\le m\le4$~\cite{georgi}. 
Of these choices only $m=1,~4$ allow
$\rho=1$, but in this case $ \sin^2\theta_W = 0.75,~ 0$,
both of which are phenomenologically uninteresting.~\footnote{These are 
the values of $ \theta_W $ at a scale $ 1/L$, and
will be modified by the renormalization 
group evolution down to the electroweak scale, but the
change needed to match the experimental data could be obtained
only if $ 1/L \gg $ several $\tev$ which, aside from
resurrecting the hierarchy problem, is inconsistent
with one of the original arguments for introducing the model.}

{\it(ii)} If $-\betbf \pm \alpbf$ is a root
then $[ E_{-\betbf} , E_{\pm\alpbf}] \propto E_{-\betbf\pm\alpbf} $,
but now $[ E_{-\betbf} , E_{-\betbf \pm\alpbf}] $ must 
vanish~\footnote{Else it would belong to $P^+$ 
and so must be $ \propto E_\alpbf$ or $ \propto E_{-\alpbf}$ which 
is impossible; \ie\  for the same reasons leading to 
(\ref{vancom}).}. In this case 
$ \betbf \cdot( \betbf \pm\alpbf) =|\betbf|^2 $ 
whence $ \sin^2\theta_W=1$
which is again phenomenologically uninteresting.

{\it Non-simple groups}
When the gauge group is not simple (\ref{star}) is
replaced by 
$ \sum_{def} g_d\vBB_{d a} \vBB_{e b} \vBB_{f c} f_{ d e f} = g_a f_{abc}  $
(and an equivalent expression for $\tilde\vBB$)
where the $g_a$ denote the gauge coupling constants taking the
same value for all indices $a$ belonging to one group factor.
These imply that if $\vBB$ maps the gauge fields of some factor 
group $G_i$ into those of another factor $G_j$, then
these groups must have the same algebras and gauge couplings.
Models where this is not trivial ($i\not=j$) have a
gauge group of the form $ G^N \times\cdots$ where the
$N$ factors of $G$ have the same couplings constants
and so have an additional permutation symmetry $\pcal$ which
is respected by $\vBB$ and $\tilde\vBB$.

Phenomenologically we must require that the low-energy gauge fields
be singlets under $\pcal$ (else the light gauge bosons
would be members of a non-trivial $\pcal$ multiplet so that the
electroweak gauge group would be of the form $ \su2^n\times\ui^l$ 
for some integers $n,l>1$). The
$\su2\times\ui$ generators will be a direct sum of generators
of the form (\ref{su2gen}, \ref{u1gen}) with one contribution
from each of the $N$ factor groups (each term
containing the same $\alpbf$ and $\hat\yy$ as a
result of the invariance under $\pcal$). Hence the crucial
expressions (\ref{abi}, \ref{lmass}) remain unchanged and
the same problems associated with $\rho$ and $\theta_W$ occur.

{\it Conclusions}
We have shown that within the gauge-Higgs unification scenario
(with neither brane gauge kinetic terms nor anomalous gauge group factors)
in 5D the phenomenological conditions (\ref{phencon})
necessarily imply a light electroweak gauge group $G_{\rm light} $
larger than $\su2\times\ui$. This general statement is 
illustrated by specific cases that have appeared in the 
literature, e.g. \cite{Scrucca:2003ra} ($\sin^2\theta_W=3/4$) and 
\cite{Haba:2004qf} (extra $U(1)$ factor in $G_{\rm light} $).
It is unlikely that GHU model with an extended $G_{\rm light} $
can be phenomenologically viable since this would require 
the Hosotani mechanism to  generate a two stage breaking, 
$G_{\rm light} \to \su2\times\ui$ at a scale $V$
by one 4D scalar mode $\Phi$, and 
$ \su2\times\ui \to \ui$ at a scale $ v \ll V$ by another mode $ \phi$.
But this hierarchy is determined by the 1-loop effective 
potential generated by all bosonic and fermionic modes,
which mix $\Phi$ and $\phi $. Then, in the absence
of fine tuning, the hierarchy $V \gg v$ cannot be maintained.

These results do not necessarily generalize to more
than 5 dimensions\cite{Csaki:2002ur}. The conditions under which
models in $\ge6 $ dimensions are phenomenologically
viable will be examined in a future publication.

\acknowledgments
This work is supported in part by the State
Committee for Scientific Research (Poland) under grant 1~P03B~078~26
in the period 2004-07, and by funds provided 
by the U.S. Department of Energy under grant No.~DE-FG03-94ER40837.

\end{document}